\documentclass[11pt]{article}
\usepackage{hyperref}
\pdfoutput=1
\begin{document}
\title{Flying with Abrupt Wing Flapping: Damselfly in Darting Flight of \\ Fluid Dynamics Videos}
\author{Chengyu Li, Haibo Dong, and Wen Zhang \\
\\\vspace{6pt} Department of Mechanical and Aerospace Engineering, \\ University of Virginia, Charlottesville, VA 22904, USA}
\maketitle

\begin{abstract}
Damselflies show abrupt, darting flight, which is the envy of aero-engineers. This amazing ability is used both to capture prey and, by males, to establish territories that can attract females. In this work, high-resolution, high-speed videos of a damselfly (Hetaerina Americana) in darting flight were obtained using a photogrammetry system. Using a 3D subdivision surface reconstruction methodology, the damselfly's wing deformation and kinematics were modeled and reconstructed from those videos.  High fidelity simulations were then carried out in order to understand vortex formation in both near-field and far-field of damselfly wings and examine the associated aerodynamic performance. A Cartesian grid based sharp interface immersed boundary solver was used to handle such flows in all their complexity.
\end{abstract}

\end{document}